\documentclass[preprint,amsmath,amssymb,superscriptaddress,notitlepage]{revtex4-1}
\usepackage{epsfig}
\usepackage{graphicx}
\usepackage{color}
\usepackage{bm}

\begin{document}

\title{Temperature and thickness dependence of tunneling anisotropic magnetoresistance in exchange-biased Py/IrMn/MgO/Ta stacks}

\author{H.~Reichlov\'a}
\affiliation{Institute of Physics ASCR, v.v.i., Cukrovarnick\'a 10, 162 53, Praha 6, Czech Republic}  
\affiliation{Faculty of Mathematics and Physics, Charles University in Prague, Ke Karlovu 3, 121 16 Prague 2, Czech Republic}
\author{V.~Nov\'ak}
\affiliation{Institute of Physics ASCR, v.v.i., Cukrovarnick\'a 10, 162 53, Praha 6, Czech Republic}   
\author{Y.~Kurosaki}
\affiliation{Hitachi Ltd., Central Research Laboratory, 1-280 Higashi-koigakubo, Kokubunju-shi, Tokyo 185-8601, Japan} 
\author{M.~Yamada}
\affiliation{Hitachi Ltd., Central Research Laboratory, 1-280 Higashi-koigakubo, Kokubunju-shi, Tokyo 185-8601, Japan}
\author{H.~Yamamoto}
\affiliation{Hitachi Ltd., Central Research Laboratory, 1-280 Higashi-koigakubo, Kokubunju-shi, Tokyo 185-8601, Japan}
\author{A.~Nishide}
\affiliation{Hitachi Ltd., Central Research Laboratory, 1-280 Higashi-koigakubo, Kokubunju-shi, Tokyo 185-8601, Japan}
\author{J.~Hayakawa}
\affiliation{Hitachi Ltd., Central Research Laboratory, 1-280 Higashi-koigakubo, Kokubunju-shi, Tokyo 185-8601, Japan}
\author{H.~Takahashi}
\affiliation{Hitachi Ltd., Central Research Laboratory, 1-280 Higashi-koigakubo, Kokubunju-shi, Tokyo 185-8601, Japan}
\author{M.~Mary\v{s}ko}
\affiliation{Institute of Physics ASCR, v.v.i., Cukrovarnick\'a 10, 162 53, Praha 6, Czech Republic}  
\author{J.~Wunderlich}
\affiliation{Institute of Physics ASCR, v.v.i., Cukrovarnick\'a 10, 162 53, Praha 6, Czech Republic} 
\affiliation{Hitachi Cambridge Laboratory, Cambridge CB3 0HE, United Kingdom}
\author{X.~Marti}
\affiliation{Institute of Physics ASCR, v.v.i., Cukrovarnick\'a 10, 162 53, Praha 6, Czech Republic} 
\author{T.~Jungwirth}
\affiliation{Institute of Physics ASCR, v.v.i., Cukrovarnick\'a 10, 162 53, Praha 6, Czech Republic} 
\affiliation{School of Physics and Astronomy, University of Nottingham, Nottingham NG7 2RD, United Kingdom}

\begin{abstract}
We investigate the thickness and temperature dependence of a series of Ni$_{0.8}$Fe$_{0.2}$/Ir$_{0.2}$Mn$_{0.8}$ bilayer samples with varying thickness ratio of the ferromagnet/antiferromagnet ($\text{t}_{FM}/ \text{t}_{AFM}$) in order to explore the exchange coupling strengths in tunneling anisotropic magnetoresistance (TAMR) devices. Specific values of $\text{t}_{FM}/ \text{t}_{AFM}$ lead to four distinct scenarios with specific electric responses to moderate magnetic fields. The characteristic dependence of the measured TAMR signal on applied voltage allows us to confirm its persistence up to room temperature despite an overlapped contribution by a thermal magnetic noise.
\end{abstract}

\maketitle
The field of spintronics is increasingly embracing antiferromagnets (AFM) as active layers in the heterostructures \cite{Jungwirth2016}. Despite being often denoted as magnetically rigid, there are a number of recent reports that describe methods to manipulate the staggered moments in AFMs in order to demonstrate basic proof of concept devices. For instance, the latest theoretical proposal, recently realized also experimentally \cite{Reichlova2015,Wadley2016}, is to run electrical currents that would torque the AFM moments at moderate current densities \cite{Zelezny2014}. The successful reorientation of AFM moments in \cite{Wadley2016}, however, requires local inversion asymmetries on the AFM spin-sublattices and therefore is not applicable to every AFM material.

Among the established experimental methods for manipulating AFM moments, the magnetic field-induced spin-flop \cite{Chu2010, Kriegner2016} is the most straightforward one which, however, typically requires high magnetic fields. \cite{Muir1971}. Another method is based on the exchange-coupling effect (ECE) which couples AFM and ferromagnetic (FM) layers at their interface. It has been successfully employed for manipulating magnetic moments from cryogenic up to room temperatures \cite{Park2011,Wang2012}.

The ECE is an important phenomenon on which many of modern spintronic devices are relying and as such has been studied extensively \cite{Nogues1999,Scholl2004,Ali2006}. Its strength and characteristics strongly depend on the sample structure giving rise to a very rich parameter space. Among the plethora of existing variables, in this paper we chose to scrutinize the role of the relative thickness of the adjacent FM and AFM layers. As in the previous publication \citep{Park2011, Marti2012}, we employ the TAMR as a methodology for electrically reading out the orientation of AFM moments \cite{Marti2012}. We further study the effect of temperature for various AFM thicknesses and we show that the choice of the relative thickness is critical to observe the TAMR at room temperature.

The devices described in this manuscript contain the stacking of layers sketched in Fig.~1a : Si/SiO$_2$/Ta(5)/ Ru(50)/Ta(5)/Ni$_{0.8}$Fe$_{0.2}$(10)/Ir$_{0.2}$Mn$_{0.8}$(t)/MgO(2.5)/Pt(10), where numbers in brackets are thicknesses in nm and the parameter t is the thickness varying in the range of 1.5-10~nm. The layers were prepared by UHV RF magnetron sputtering on a thermally oxidized silicon substrate. The layers were grown under magnetic field of 5~mT. The multilayers were annealed at 350$^{\circ}$C for 1 hour in magnetic field 0.4~T applied along the same direction as during the growth. Several tunnel junctions were fabricated from each wafer, the dimensions were 3x6~$\mu$m$^2$ and 5x10~$\mu$m$^2$. Within the stack, the tunnel junction comprises an AFM (IrMn) layer, a MgO barrier, and a non-magnetic layer. The AFM layer is adjacent to a FM layer (NiFe) forming an exchange-bias system. The confirmation of the interface coupling is demonstrated by the systematically broadened and/or shifted magnetization loops as observed by the superconducting quantum interference device (SQUID) magnetometry (Fig.~1b). 

The magnetoresistance is measured under a constant voltage (V) applied across the tunneling barrier and with the magnetic field applied in the plane of the sample. In this experimental set-up, the tunneling current senses only the changes in the AFM magnetic configuration and responds to changes in the FM layer only indirectly if the latter is capable of modifying the AFM configuration. Further experimental details are described elsewhere \cite{Park2011, Marti2012}.

\begin{figure}[h]
\hspace*{0cm}\epsfig{width=0.8\columnwidth,angle=0,file=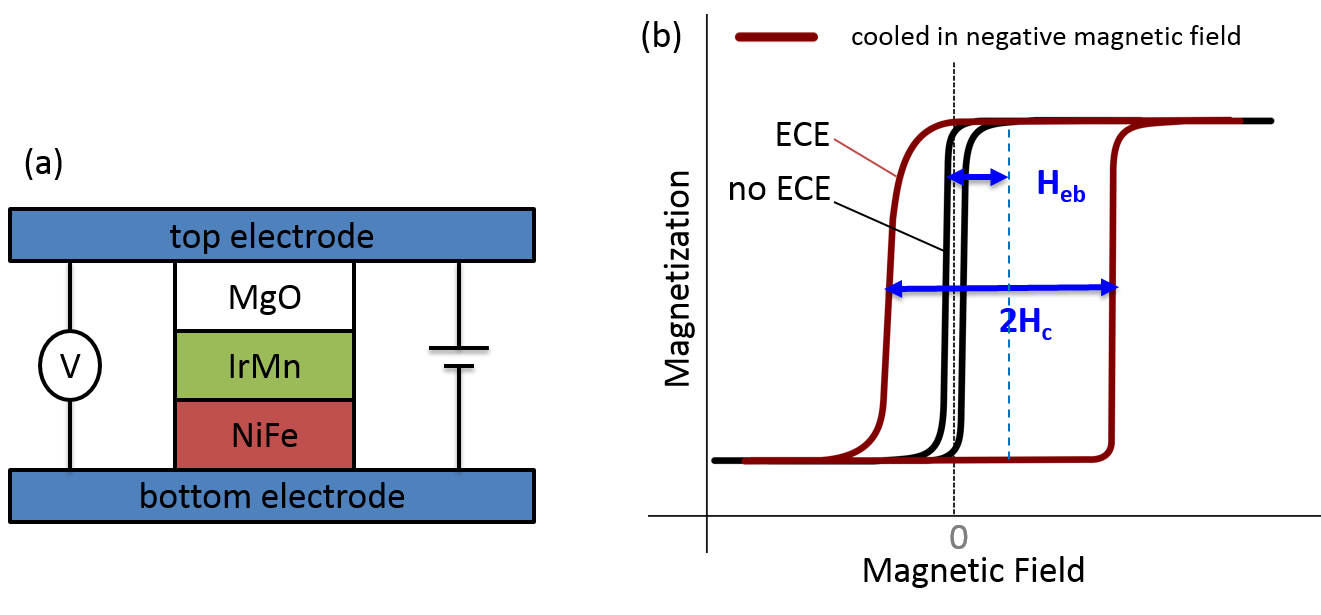}
\caption{(a) Schematic illustration of a TAMR structure with used stack of layers. Note, that the magnetoresistance of our tunnel structure is caused by the IrMn/MgO interface. (b) Schematic illustration of the exchange coupling effect and its two characteristics - exchange bias $\text{H}_{eb}$ and by exchange broadening (the increased coercive field $\text{H}_{c}$). }
\label{f1}
\end{figure}

The numerous models of the exchange coupling effect generally agree on the critical role of the relative FM/AFM thickness as a key parameter defining the ECE and that both the broadening (2 $\text{H}_{c}$) and the shift ($\text{H}_{eb}$) of the magnetization loop are sensitive indicators of changes in the ECE \cite{Nogues1999,Ali2003,Moritz2012}. Temperature is another important parameter for the ECE which we vary in our measurements. We used SQUID magnetometry to study a set of FM/AFM bilayers with continuously increasing the thickness of the AFM in order to detect an optimal the FM/AFM ratio for the ECE and the TAMR. 

The systematic mapping of the ECE is shown in Fig. 2a and 2b displaying $\text{H}_{eb}$ and $\text{H}_{c}$, respectively. The data show that $\text{H}_{c}$ and $\text{H}_{eb}$ have distinct dependencies on the temperature and on the IrMn thickness. Also, the critical temperatures where the exchange broadening and exchange shift vanish do not coincide. According to the anisotropy energy balance \cite{Nogues1999}, the observation of the broadening can be ascribed to the AFM moments fully rotating along with the FM moments. On the other hand, shifts of the hysteresis loop would correspond to a rigid AFM pinning the FM layer. The simultaneous existence of both features indicates intermediate cases with partial pinning and partial reorientation of the AFM moments. By presenting the data in a simplified form showing only the presence/absence of non-zero $\text{H}_{eb}$ or $\text{H}_{c}$, four distinct regions can be identified in the phase diagram (Fig. 2c).

\begin{figure}[h]
\hspace*{0cm}\epsfig{width=1\columnwidth,angle=0,file=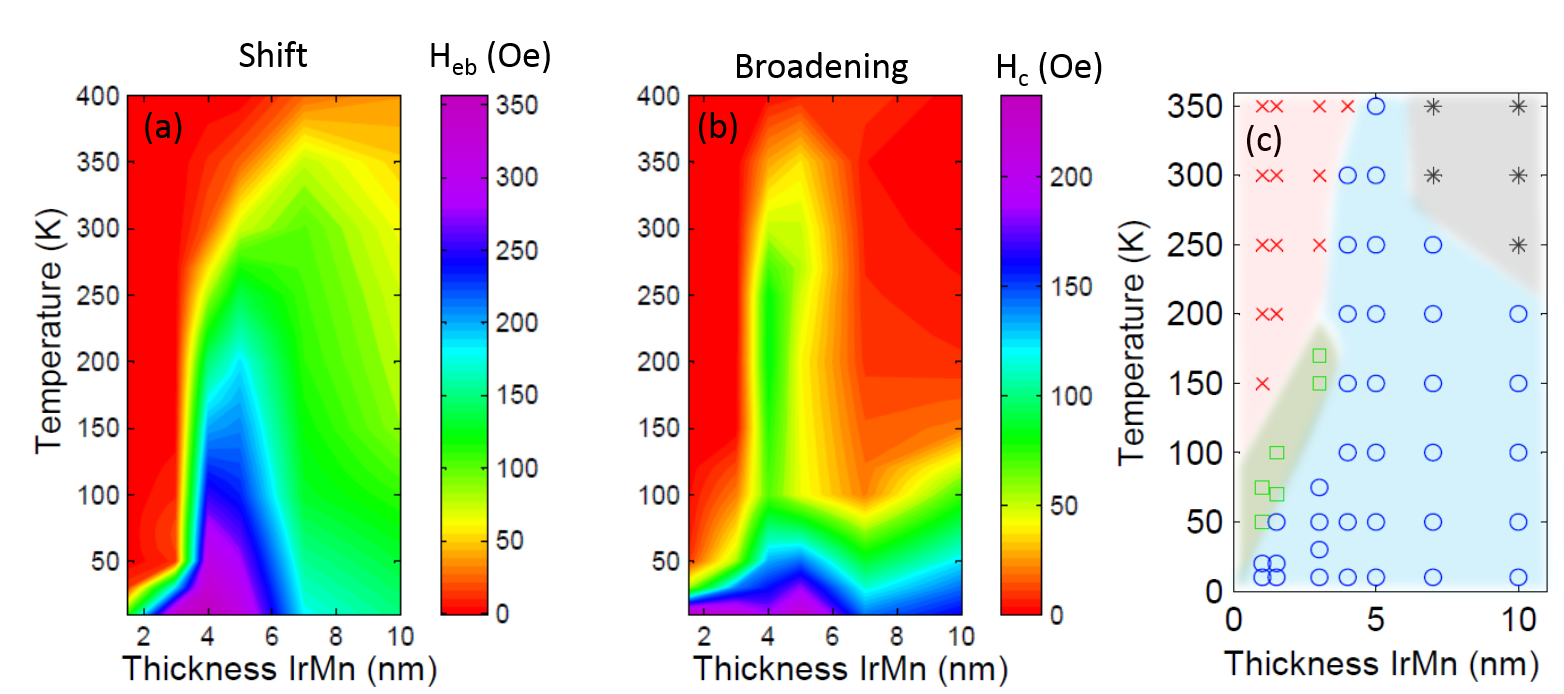}
\caption{(a) $\text{H}_{eb}$ as a function of IrMn thickness and temperature. (b) $H_{c}$ as a function of IrMn thickness and temperature. (c) Phase diagram of IrMn thickness and temperature showing four distinct regions depending on if the AFM coupled moments can be manipulated. Red crosses correspond to the region where no ECE is observed. Black stars correspond to the region where AFM is rigid and cannot be manipulated. Green squares correspond to the region when AFM has low anisotropy and fully follows the FM layer. Blue circles correspond to the region where AFM coupled moments are dragged by FM and partially rotated.}
\label{f2}
\end{figure}

The first phase (red crosses) represents the scenario, in which no ECE is observed. In the studied temperature range this occurs in thin AFMs in which the Neel temperature is reduced due to the reduced thickness \cite{Molina-Ruiz2011}. Consequently, the blocking temperature of the ECE is also lowered. The second phase (green  squares) represents the phase where only $\text{H}_{c}$ has a non-zero value and no $\text{H}_{eb}$ is detected. This corresponds to a weak anisotropy of the AFM layer, which is, however, coupled to the FM so that the two layers follow together the external magnetic field $\text{H}_{ext}$. The third phase (black stars) represents the opposite scenario to the second phase: the anisotropy of the AFM layer is high and the moments cannot be manipulated by the FM. The AFM is, however, still coupled to the FM which increases $\text{H}_{eb}$ of the magnetization loop, with no increase in $\text{H}_{c}$. The last phase (blue circles) represents the scenario in which the AFM is coupled to the FM and it follows the FM whose moments are rotated by $\text{H}_{ext}$. The AFM moments, however, rotate only partially \cite{Scholl2004}, unlike the second scenario, in which the AFM moments are rotated fully.

Having described the four magnetic scenarios (Fig.~2c), we now turn to the consequences of such scenarios for the tunneling transport, which will be illustrated on one particular thickness of IrMn (3~nm) measured at different temperatures. First, the absence of both $\text{H}_{c}$ and $\text{H}_{eb}$ suggests a weak interaction among the FM and AFM layers. In this case, $\text{H}_{ext}$ will align the FM moments but the ECE between FM and AF is not strong enough to reorient the AF moments. Then the tunneling current remains unchanged because the AFM moment orientation remains unchanged. This scenario corresponds to the constant resistance value observed in Fig.~3a (3~nm, 200~K). The second scenario, which has been successfully exploited in FM-based data storage technology, corresponds to the case of a rigid AFM (thickness $\sim$ 10~nm) pinning the FM layer. In this case, $\text{H}_{ext}$ is not strong enough to be able to manipulate the rigid AFM and only manipulates the FM moments. The structure of our TAMR device (shown in Fig.~1a) exploits the AFM layer as an active element of the tunnel structure and therefore the TAMR has no direct dependence on the orientation of the magnetic moments in the FM layer. Consequently $\text{H}_{ext}$ has no impact on the measured TAMR. The third case corresponds to a soft AFM layer rotating fully with the FM (Fig.~3b). This is an important case as it corresponds to the largest change in the angle of the AFM moments when reversing the FM. Translated into the TAMR signal, our data displays a significant peak-shaped feature with signal-to-noise ratio of ~10 in a range of 5000~$\Omega \mu m^2$. This figure is a representative upper bound for the TAMR magnitude in a specific device at the given temperature. We note that the entire rotation of the AFM moments occurs within a moderate $\sim$ 500~Oe fields compared to the typical spin-flop fields of the order of several to tens of Tesla. The last scenario corresponds to a partial rotation of the AFM by the FM. Upon the FM reversal the AFM moments can be stabilized in two metastable configurations corresponding to distinct tunneling magnetoresistance signals. This is the case of the data shown in the panels Fig.~3c. The order of magnitude of the separation of the two TAMR states is consistent with the bounds set by the fully-rotating scenario in Fig.~3b when considering the difference in the temperature between the two measurements and the expected reduction of the TAMR signal with increasing temperature. Notably, the magnetic fields required for switching an AFM among two stable configurations is still of the order of $\sim$ 500~Oe. The hysteretic behavior shown in Fig.~3c reflects the $\text{H}_{c}$ of the exchange coupled NiFe/IrMn bilayer.

\begin{figure}[h]
\hspace*{0cm}\epsfig{width=1\columnwidth,angle=0,file=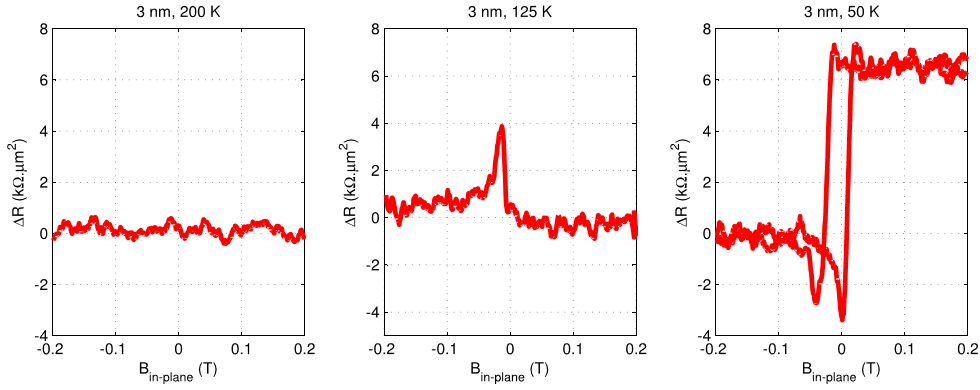}
\caption{ TAMR signal of sample with 3~nm IrMn measured at 50~K (a),  125~K (b) and 200~K (c). Magnetic field is applied in the plane of the sample. Applied voltage V~=~20~mV.}
\label{f3}
\end{figure}

The correlation of the phase diagram in Fig.~2c and the TAMR in Fig.~3 allows us to search for the highest temperature at which the TAMR is still present. According to the observations described so far, the 4/10 relative thickness of the AFM vs FM layer gives finite $\text{H}_{c}$ and $\text{H}_{eb}$ at room temperature. As can be seen in the phase diagram (Fig.~2c) the optimal thickness of IrMn is 4-5~nm, in which both $\text{H}_{c}$ and $\text{H}_{eb}$ persist up to room temperature. However, the inspection of the TAMR data for this thickness reveals a more complex scenario.

Although the expected TAMR signal is measured at lower temperatures (an example of the measurement at 100~K is shown in Fig.~4a), at higher temperatures a different behavior is observed, as can be seen in Fig.~4b. A statistical analysis of the data in Fig.~4b reveals a stochastic switching between well defined states, consistent with the data at lower temperatures (Fig.~4a). This intriguing coincidence points to a TAMR origin of the data in Fig.~4b overlapped with an additional effect. The two states cannot be stabilized by applying magnetic field, instead a random switching  is observed. Because this behavior is present only at elevated temperatures (starting at $\sim$250~K) we conjecture that the additional effect is a thermal fluctuation causing random switching between the two TAMR states.

\begin{figure}[h]
\hspace*{0cm}\epsfig{width=1\columnwidth,angle=0,file=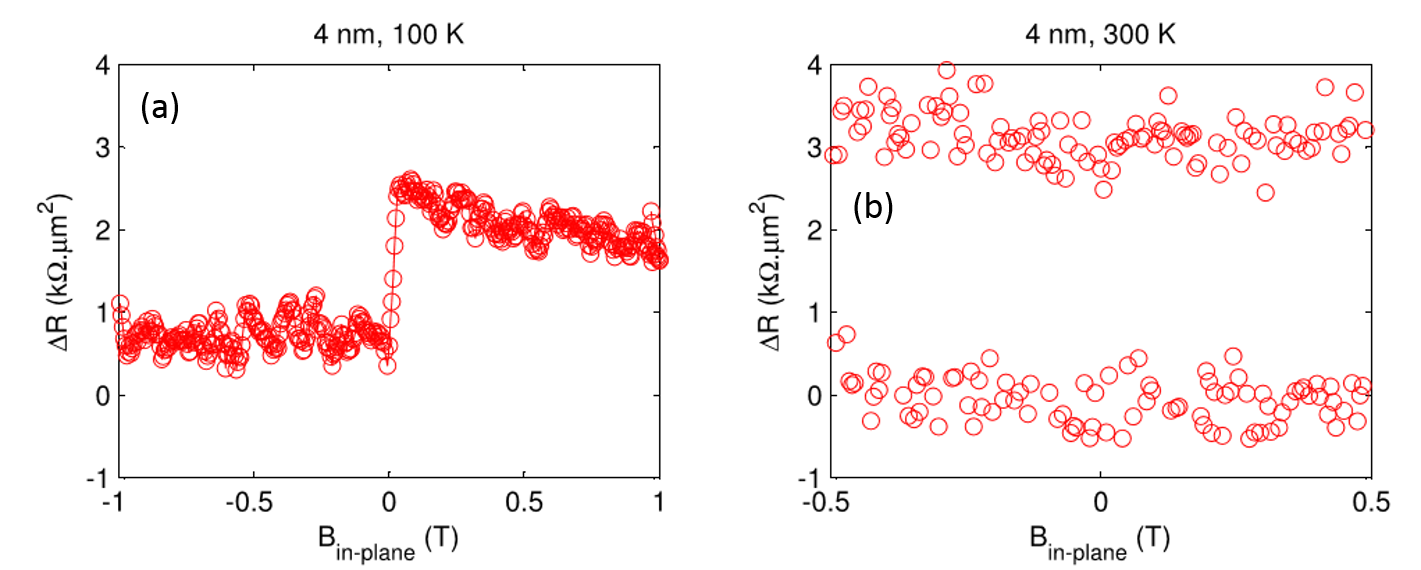}
\caption{TAMR signal of sample with 4~nm IrMn measured at 100~K (a) and 300~K (b). Magnetic field is applied in the plane of the sample. Applied voltage V~=~100~mV.}
\label{f4}
\end{figure}

To confirm that the two states in Fig.~4b are of the TAMR origin we present an additional feature. In Fig.~5a, we show a typical voltage dependence of a TAMR device with 3~nm IrMn at 75~K. The voltage dependence was obtained by subtracting two R(V) curves measured under different magnetic field. The TAMR is then expressed as a percentage of the total resistance. The figure shows the asymmetry with respect to zero with the maximal TAMR signal at a negative voltage. Similar voltage dependence of the TAMR is typical for all measured devices, as illustrated in Fig.~5b and Fig.~5c where data measured at 100~K for two different thicknesses of IrMn are presented. Note that the voltage at which the TAMR reaches its maximum varies for different samples due to the different resistivity of different tunnel junctions. In Fig.~4b we have presented switching between two distinct states corresponding to two different values of TAMR, however, the two states can not be stabilized by an external magnetic field, as in Fig.~4a. Nevertheless, if the difference between the two states in Fig.~4b is of the TAMR origin, we would expect to observe a voltage dependence of this difference similar to TAMR(V) dependencies presented in Figs.~5a-c. In Fig.~5d, we show the voltage dependence of the difference between two states presented in Fig.~4b. By subtracting two R(V) curves (note that in this case it does not matter if $\text{H}_{ext}$ was applied or not), we obtain the characteristic TAMR(V) dependence evidencing the asymmetry around zero. Therefore, we confirmed the TAMR origin of the signal observed at 300~K.

\begin{figure}[h]
\hspace*{0cm}\epsfig{width=0.8\columnwidth,angle=0,file=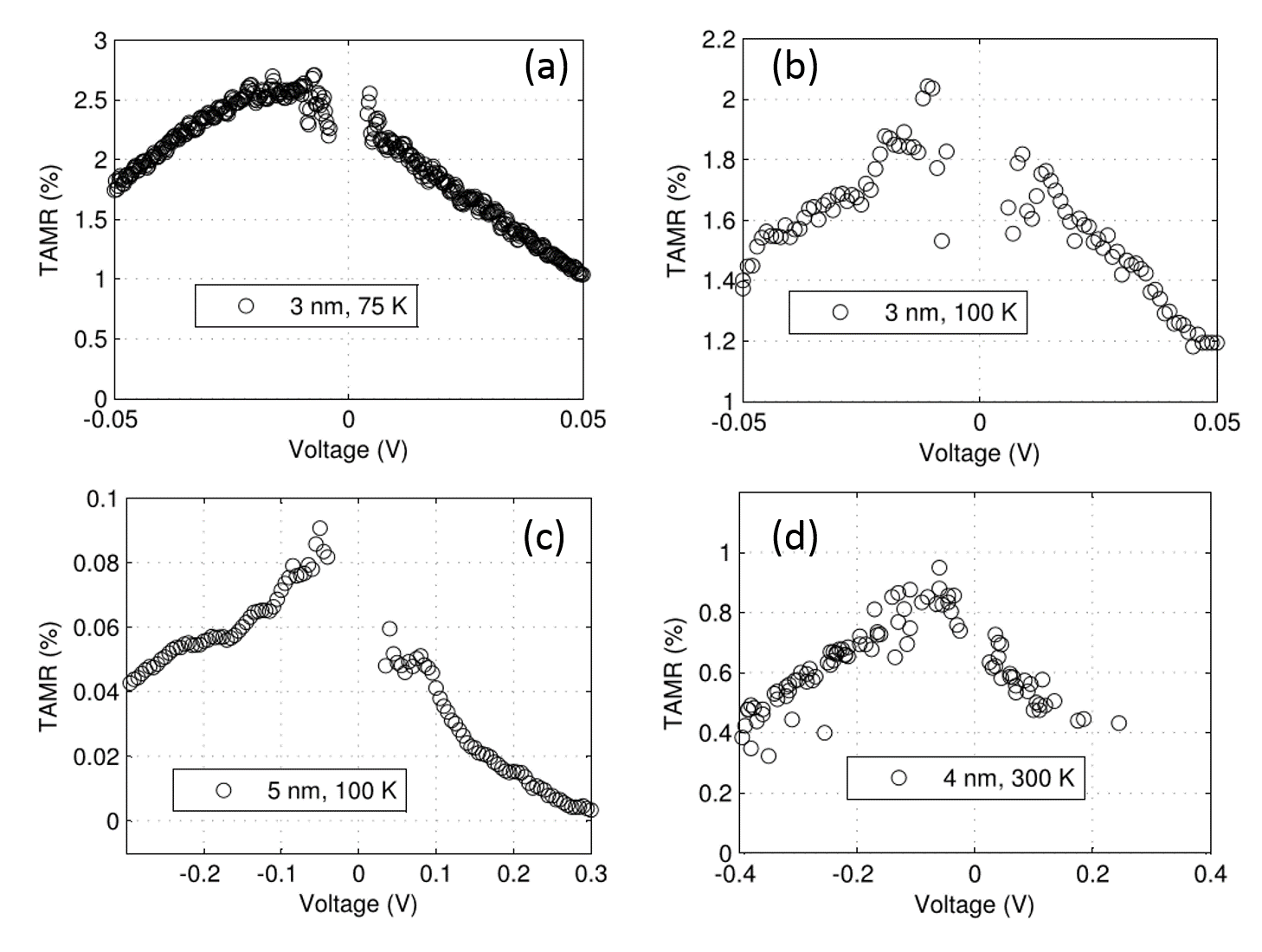}
\caption{TAMR voltage dependence for sample with IrMn 3~nm at 75~K (a), IrMn 3~nm at 100~K (b), IrMn 5~nm at 100~K (c) and IrMn 4~nm at 300~K. In panels (a), (b), (c) the TAMR voltage dependence is achieved by subtracting two R(V) curves measured under $\text{H}_{ext}$ with opposite polarity. In the panel (d)  $\text{H}_{ext}$ has no effect on the measured R(V) curve, therefore, by subtracting any two R(V) curves we obtain the presented voltage dependence.}
\label{f5}
\end{figure}

\pagebreak

In summary, we have studied systematically the effect of the thickness of the IrMn on the exchange coupling at different temperatures by SQUID magnetometry. We have identified four different regions with specific exchange coupling features. Having confirmed the correspondence of the magnetic phase diagram and the TAMR data, we found an optimal AFM/FM thickness ratio of 4/10 for observing the TAMR at room temperature. This is indeed experimentally confirmed. However, the TAMR signal is not stable in our devices as the AFM appears to fluctuate stochastically between the distinct TAMR states at room temperature.

We acknowledge financial support by Ministry of Education of the Czech Republic grant No. LM2015087. EU
ERC Synergy Grant No. 610115, the Grant Agency of the Czech Republic Grant No. 14-37427G.

\bibliographystyle{ieeetr} 
%\bibliography{references}

\end{document}